\documentclass[manuscript]{acmart}

\usepackage{booktabs}
\usepackage{multirow}
\usepackage{subcaption} 
\usepackage{arydshln}
\usepackage{soul}
\usepackage{enumitem}

\AtBeginDocument{%
  \providecommand\BibTeX{{%
    \normalfont B\kern-0.5em{\scshape i\kern-0.25em b}\kern-0.8em\TeX}}}

\setcopyright{acmcopyright}
\copyrightyear{2018}
\acmYear{2021}
\acmDOI{10.1145/1122445.1122456}

\acmConference[RecSys '21]{RecSys '21: ACM Conference Series on Recommender Systems}{September 27--October 05, 2021}{Amsterdam, NL}
\acmBooktitle{RecSys '21: ACM Conference Series on Recommender Systems,
  September 27--October 05, 2021, Amsterdam, NL}
\acmPrice{15.00}
\acmISBN{978-1-4503-XXXX-X/18/06}

\definecolor{lcs}{RGB}{153,50,204}
\definecolor{tabd}{RGB}{255,70,0}

\newcommand{\colored}[1]{{\bf \color{tabd} #1}}

\begin{document}


\title{Topic Model Robustness to Automatic Speech Recognition Errors in Podcast Transcripts}


\author{Raluca Alexandra Fetic}
\authornotemark[1]
\affiliation{%
  \institution{Podimo}
  \streetaddress{Gasværksvej 16}
  \city{Copenhagen V}
  \country{Denmark}
  \postcode{DK-1654}
}
\author{Mikkel Jordahn}
\authornotemark[1]
\affiliation{%
  \institution{DTU Compute, Technical University of Denmark}
  \streetaddress{Richard Petersens Plds.\ B324}
  \city{Kongens Lyngby}
  \country{Denmark}
  \postcode{DK-2800}
}
\author{Lucas Chaves Lima}
\authornotemark[1]
\affiliation{%
  \institution{Podimo}
  \streetaddress{Gasværksvej 16}
  \city{Copenhagen V}
  \country{Denmark}
  \postcode{DK-1654}
}
\author{Rasmus Arpe Fogh Egebæk}
\affiliation{%
  \institution{DTU Compute, Technical University of Denmark}
  \streetaddress{Richard Petersens Plds.\ B324}
  \city{Kongens Lyngby}
  \country{Denmark}
  \postcode{DK-2800}
}
\author{Martin Carsten Nielsen}
\affiliation{%
  \institution{DTU Compute, Technical University of Denmark}
  \streetaddress{Richard Petersens Plds.\ B324}
  \city{Kongens Lyngby}
  \country{Denmark}
  \postcode{DK-2800}
}
\author{Benjamin Biering}
\affiliation{%
  \institution{Podimo}
  \streetaddress{Gasværksvej 16}
  \city{Copenhagen V}
  \country{Denmark}
  \postcode{DK-1654}
}
\author{Lars Kai Hansen}
\affiliation{%
  \institution{DTU Compute, Technical University of Denmark}
  \streetaddress{Richard Petersens Plds.\ B324}
  \city{Kongens Lyngby}
  \country{Denmark}
  \postcode{DK-2800}
}

\renewcommand{\shortauthors}{Jordahn, Fetic and Chaves, et al.}

\begin{abstract}
For a multilingual podcast streaming service, it is critical to be able to deliver relevant content to all users independent of language. Podcast content relevance is conventionally determined using various metadata sources. However, with the increasing quality of speech recognition in many languages, utilizing automatic transcriptions to provide better content recommendations becomes possible. In this work, we explore the robustness of a Latent Dirichlet Allocation topic model when applied to transcripts created by an automatic speech recognition engine. Specifically, we explore how increasing transcription noise influences topics obtained from transcriptions in Danish; a low resource language. First, we observe a baseline of cosine similarity scores between topic vectors from automatic transcriptions and the descriptions of the podcasts written by the podcast creators. We then observe how the cosine similarities between topic vectors decrease as transcription noise increases and conclude that even when automatic speech recognition transcripts are erroneous, it is still possible to obtain high-quality topic vectors.

\end{abstract}

\ccsdesc[500]{Information systems~Information retrieval}
\ccsdesc[500]{Computing methodologies~Natural language processing}

\keywords{Podcasts, Automatic Speech Recognition, Topic modeling, Recommendation Systems}

\maketitle

\section{Introduction}

Podcasts have become an increasingly popular audio format in recent years. Podcasts encompass a variety of on-demand audio, such as radio, news, and entertainment in the form of informal discussions, interviews, or even narrated content similar to audiobooks, found in several different categories. Despite the growth in popularity of podcasts, an open challenge is how to find a new podcast to listen to. Research in Podcast recommendation has not yet been able to follow up and provide users with efficient and high-quality recommendations which are key to ensuring a high quality of streaming services~\cite{jones2021current}. Podcast recommendation is a challenging task considering the very large amount of podcast episodes that lack metadata on both podcast and episode levels. Previous work has shown that transcription-based topic modeling plays a crucial role in podcast recommendation, as users tend to focus on the topics of the podcasts instead of podcast audio style \cite{molgaard2007castsearch, morethanwords}.

Topic modeling techniques such as Latent Dirichlet Allocation (LDA) \cite{LDA2002,LDA2003} and Probabilistic Latent Semantic Indexing (PLSI) \cite{pLSI1999} are widely used to discover topics over high-quality texts (e.g. news, blog posts, etc.). As podcasts are usually not scripted, a transcript generated from an Automatic Speech Recognition (ASR) system is necessary to perform topic modeling. Podcasts represent a particularly challenging audio format for speech recognition systems as a large number of artifacts are commonly present. Some of the challenges are that multiple speakers and speech overlap, background music and jingles, audio effects, and "real-life" recording conditions (e.g. background noise), which makes the ASR-generated transcripts from podcasts prone to errors. 

While ASR systems have been widely explored and developed for high-resource languages such as English, there is a significant lack in many low-resource languages\footnote{By low resource language, we refer to a smaller language with a limited amount of training data for NLP tasks such as speech recognition.}. As such the systems available in such setups can be expected to produce more errors than their high-resource counterparts. This motivates research for downstream tasks in the low-resource setup to ensure multilingual application. In this paper, we study the robustness of an LDA topic model used on podcast transcriptions generated by an ASR engine in Danish; a low resource language. We utilize podcast episodes with author descriptions and assume that these descriptions contain a good representation of topics. Hence, a high similarity between topic vectors from author descriptions and automated transcripts also indicates a good representation of topics in the automated transcript. To evaluate the robustness of the topic model, we first construct a baseline by computing a cosine similarity between the topic vector representations of the author descriptions and automated transcripts. We then introduce noise to the automated transcripts, controlled by a variable noise injection rate, determining how often words should be replaced, and observe how the cosine similarity changes. We experiment with two types of noise; simulated ASR noise sampled from a conditional distribution derived from a transcription error dataset (see Section~\ref{ss:noiseinjection}), and words sampled uniformly from the topic model vocabulary for reference. Empirically, over a dataset of 587 episodes from 24 Danish podcasts, we find that the topic model is much more robust to simulated ASR noise than it is to noise from a uniform distribution. We present evidence that the LDA topic model is robust and captures an informative representation of topics, even in the face of imperfect transcriptions.







The remainder of this paper is structured as follows. In Section~\ref{s:relatedwork}, we present the necessary background on topic modeling and robustness of downstream NLP systems against ASR errors. Section~\ref{s:podcasttopicmodeling} defines the problem and details the methods used by each component to test the podcast topic modeling robustness. Section~\ref{s:experimentalsetup} presents the experimental setup. In Section \ref{s:resultsanddiscussion}, we present and discuss the results. Lastly, in Section~\ref{s:conclusion}, we conclude the paper and propose future research directions.

\section{Background}
\label{s:relatedwork}
\subsection{Automatic Speech Recognition}
\label{ss:automaticspeechrecognition}
Converting speech in audio to raw text is done using an Automatic Speech Recognition (ASR) system. Speech recognition systems have drastically improved during recent years with advancements such as various data augmentation techniques \cite{park2019specaugment}, pre-training procedures on unlabelled speech data \cite{baevski2020wav2vec} and noisy student training \cite{park2020improved}. State of the art performance on the common English benchmark dataset LibriSpeech \cite{panayotov2015librispeech} is as low as 1.4\% and 3.3\% of Word Error Rate (WER) on the test-clean and the test-other partitions, respectively \cite{zhang2020pushing}.
Labeled speech recognition training data is highly accessible in English \cite{panayotov2015librispeech, kahn2020libri} but less so in many other languages. Even with crowd-sourcing data initiatives such as CommonVoice from Mozilla \cite{ardila2019common}, the recognition performance gap between low and high resource languages remains quite high.\footnote{\url{https://paperswithcode.com/dataset/common-voice}} Pre-training speech recognition models with cross-lingual data has helped bridge the gap significantly \cite{conneau2020unsupervised}. However, transcripts from ASR systems for low resource languages are likely to be error-prone, especially for complex audio data such as podcasts. 
\subsection{Topic Modeling and Evaluation}
\label{ss:tmae}
Topic models are used to explore and structure a large set of documents according to latent semantic content. To improve downstream tasks such as search and recommendation of podcasts, a promising method is to utilize a topic model to extract the relevant topics of the podcast and hence enhance the podcast representations \cite{molgaard2007castsearch, morethanwords}. Topic modeling has been extensively studied, and various approaches exist. For instance, Latent Semantic Indexing (LSI)~\cite{LSA1990} uses Singular Value Decomposition (SVD) or Non-negative Matrix Factorization (NMF)~\cite{NMF1999} on a term by document matrix to construct a latent space representation, which can be queried for comparison of documents. An extension of LSI, Probabilistic Latent Semantic Indexing (PLSI), models topics as distributions over words, and documents as a probabilistic mixture of those topics~\cite{pLSI1999}. A similar, and very popular, approach is known as Latent Dirichlet Allocation (LDA). LDA differs from PLSI by utilizing prior Dirichlet distributions to model the topic-word distributions making it more robust to unseen data~\cite{LDA2002,dir2003}. Numerous extensions of the LDA model have been studied. One such extension, the Correlated Topic Model (CTM)~\cite{blei2006ctm}, explores the correlations among topics generated by the LDA model. Other extensions include Collective LDA \cite{collectiveLDA}, combining multiple corpora during the training of the models, and approaches that explore the influence of the age of the documents in the topics~\cite{LDAtime2006,10.1145/1281192.1281249}.


The quality of a topic model can be evaluated in different ways. A common practice is to evaluate a trained model in terms of perplexity~\cite{wallach2009evaluation} or topic coherence. Topic coherence, as opposed to a perplexity, is more similar to how humans judge the quality of topics \cite{autoeval2010}. Examples of topic coherence measures include UCI-coherence \cite{autoeval2}, $U_{mass}$ coherence~\cite{semanticcoherence2011}, and coherence based on word embeddings \cite{wordembeddingcoherence2016}.

\subsection{Robustness to Noise}
Downstream NLP tasks on ASR transcripts need to be robust to noise due to the commonality of transcription errors. Robustness to noise is often evaluated by constructing a baseline result with clean text and then injecting varying degrees of noise to the text and examining how the result changes ~\cite{topicstability, belinkov2017synthetic}. To investigate the effects of noise, it is necessary to select different ways of injecting plausible ASR noise into transcripts. A recent study explored the feasibility of improving the robustness of speech-enabled systems with three methods of noise~\cite{Cui2021-eo}; rule-based substitution which randomly substitutes a candidate word with a phonetically similar one, statistic-based confusion substitution which samples replacement words from a pre-constructed ASR confusion matrix and finally, model-based substitution utilizing a generative GPT model to directly produce ASR-like text. Another study investigated the stability of topics over noisy sources, by testing for topic model agreements~\cite{GreeneOC14}, after subjecting the training data to insertion of frequent words, deletion and rule-based phonetic substitution errors \cite{topicstability}.

The robustness of a downstream task varies a lot depending on the specific task and the type of noise. For instance, topic modeling has previously been shown to be robust to deletion of random words, whereas the insertion of new words and phonetic substitution errors has a larger negative impact on topic stability \cite{topicstability}. Another relevant downstream task, neural machine translation with character-based models, has shown to struggle even with small perturbations to the input data \cite{belinkov2017synthetic}.

\section{Methods}
\label{s:podcasttopicmodeling}
\subsection{Transcript Generation}\label{sec:transcription_engine}
We produce podcast transcripts by parsing podcast audio through a danish transcription system developed at the Technical University of Denmark (DTU) as part of the Danspeech project.\footnote{https://danspeech.github.io/danspeech/html/index.html} The system is based on the wav2vec 2.0 framework \cite{wav2vec2}. The model was pre-trained using approximately 945 hours of podcast episodes and 400 hours of audiobooks, and fine-tuned using the Connectionist Temporal Classification \cite{graves2006connectionist} (CTC) loss function on 200 hours of labeled data from the Nordisk Språkteknologi (NST) danish training dataset\footnote{\url{https://www.nb.no/sprakbanken/ressurskatalog/oai-nb-no-sbr-55/}} and 267 hours of aligned audiobook data. The Fairseq library \cite{ott2019fairseq} was used for both the pre-training procedure as well as fine-tuning. 
During inference, the ASR engine performs prefix beam search \cite{hannun2014first} with an open-source danish 3-gram language model\footnote{\url{https://danspeech.github.io/danspeech/html/lms.html\#danspeech.language_models.DSLWikiLeipzig3gram}} when decoding the probabilities emitted from the wav2vec 2.0 model.
\subsection{Noise Injection}
\label{ss:noiseinjection}
We inject noise into the ASR transcripts by means of a noise injection rate parameter, $\beta$, that determines the frequency at which we substitute words in the transcript. More specifically, for each word in a given transcript we independently decide if a substitution should take place with probability $\beta$. Examples of substitutions at various levels of $\beta$ are presented in Table \ref{tab:transcriptions}.

\begin{table}[!h]
\centering
\small
\caption{Examples of how a transcription changes with automatic speech recognition statistics-based substitutions at varying levels of the noise injection rate parameter $\beta$.}
\label{tab:transcriptions}
\begin{tabular}{@{}c|p{6.5cm}|p{7.2cm}}
\toprule
\textbf{$\beta$} & \multicolumn{1}{c|}{\textbf{Transcription}} & \textbf{Description} \\ \midrule
0 & historien er rig på spændende fortællinger om drama krig voldelig politiske omvæltninger og fyldt med mystik hemmeligheder og fascinerende menneskeskæbner & \multirow{4}{6.9cm}{Historien er rig på spændende fortællinger om drama, krig, voldelige politiske omvæltninger og er fyldt med mystik, hemmeligheder og fascinerende menneskeskæbner. \{...\}\\
\vspace{0.15cm}
Episode 4: Røde agenter\\
Revolutionen i Rusland i 1917 blev startskuddet til en international politisk kamp for at udbrede socialismen til hele verden \{ ... \} I programmet medvirker historikerne Niels Erik Rosenfeldt og Morten Møller.}  \\  \cdashline{1-2}
0.3 & \colored{kane} er rig på \colored{så} fortællinger om drama krig \st{\textbf{-------}} politiske omvæltninger \colored{har} fyldt \colored{ved} mystik \colored{er} og \colored{fascineren} menneskeskæbner &  \\ \cdashline{1-2}
1.0 & \colored{historie har rik p så fortælling er \st{\textbf{-------}} strama til \st{\textbf{-------}} \st{\textbf{-------}} \st{\textbf{-------}} er film som \st{\textbf{-------}} er var fascineren \st{\textbf{-------}}} &  \\ \bottomrule
\end{tabular}

\end{table}

When a word is selected for substitution we apply one of two methods for the substitution. The first method samples a word uniformly at random from the topic model vocabulary, and the second method uses a statistics-based confusion matrix approach (see Section~\ref{ss:statistic}). When performing statistics-based confusion substitutions, we sample replacement words $\bar{w}$ from a conditional error distribution with the candidate probability given as
\begin{equation}
    \label{noise:eq:1}
    P(\bar{w}|w) = \frac{W(\bar{w})}{\sum_{\bar{w} \in V(w)} W(\bar{w})}.
\end{equation}
Here $W(\bar{w})$ denotes the weighting of a candidate word and $V(w)$ denotes the candidate set for word $w$. 
\subsection{Topic Modeling and Document Vector Representation Similarity}
\label{sec:3:1}
%

To produce topic representations we use a general-purpose danish topic model trained on an external, multi-domain text corpus using LDA. Our motivation for doing so is two-fold; most prominently we choose to rely on an external text corpus because the topic modeling process requires large amounts of textual data, which is something that is not readily available to us in a pure podcast domain setup. Secondly, we choose LDA as the modeling framework due to its inherent probabilistic approach of representing topics, which have been shown to be robust to unseen data. 


The LDA model can be seen as a probabilistic function $\gamma(d) \mapsto \boldsymbol{d}_{T}$, which maps a given document $d$ to a topic vector $\boldsymbol{d}_{T} = [p_{t_{1}}, .., p_{t_{|T|}}]$,  where $p_{t}$ represents the probability of each topic $t$ for the document $d$. Thus to compute the similarity of topics present in a pair of documents, $d_1$ and $d_2$, we parse each document through the topic model and compute the Similarity between the resulting document-level topic vectors as follows,

\begin{equation}
    {\rm Similarity}(d_{1},d_{2}) = {\rm cos} (\gamma(d_{1}), \gamma(d_{2}))
\end{equation}
where cos is the cosine similarity. When comparing two ordered sets of document pairs, $(S_1, S_2)$, we denote the the average similarity between the sets as the CorpusSimilarity (CS) computed as,

\begin{equation}
    {\rm CS}(S_1,S_2) = \frac{1}{|S_1|}\sum_i\rm{Similarity} (S_1(i), S_2(i))
\end{equation}
where $S_k(i)$ is document $d_i$ in set $S_k$.

\section{Experiments}
\label{s:experimentalsetup}
\subsection{Podcast Dataset}
The podcast dataset we use to investigate the robustness of topic modeling consists of 587 episodes from 24 podcasts shows in Danish. The 24 podcasts shows belong to 8 categories, assigned by the content creators, such as  ``Culture \& leisure'',  ``Health \& personal development'',  ``History \& religion'' and ``True crime \& mysteries''. The podcast shows are all single speaker with a limited amount of audio artifacts such as background music and jingles. We extend the descriptions with the episode title, podcast title, podcast description, and podcast category. We only include episodes with a high-quality author description, defining high-quality descriptions as any description-transcription pairings that have an initial cosine similarity above $0.5$. The number of episodes for each podcast included in the dataset is presented in Figure \ref{fig:experiments:data} in the Appendix, in which it can be seen that the distribution of episodes across the shows is imbalanced, with some podcasts containing the majority of the episodes. We produce ASR transcripts for all the episodes by leveraging the ASR engine described in Section~\ref{sec:transcription_engine}.
%
%
%
\subsection{Statistics Based Automatic Speech Recognition Noise}
\label{ss:statistic}
To create the word-level conditional ASR error distributions used for statistics-based confusion substitution (see Section \ref{ss:noiseinjection}), we first construct a dataset containing clean text and ASR transcription pairs. We use the same ASR engine as described in Section \ref{sec:transcription_engine} with the exception that the wav2vec 2.0 model was only fine-tuned on the training dataset from NST. The data consists of approximately 77 hours of data from the NST test dataset and 267 hours of audiobook data resulting in 229,499 pairs of reference-transcript pairs. The probability of a candidate word for a given word is then obtained by counting the frequency at which the word is wrongfully transcribed as the candidate word, normalized by the number of candidates as shown in Equation \eqref{noise:eq:1}. Examples of how candidate distributions may look for specific words are presented in Table~\ref{tab:nogensinde} and Table~\ref{tab:lavet}, where the \textit{random} candidate is a grouping of many potential errors with very low probability. We can see from the tables that typical ASR errors tend to retain semantic meaning to some degree which gives rise to the hypothesis that topic models are likely to be robust to ASR noise. If an unknown word is encountered then it is simply deleted. Across the transcripts of the podcasts, unknown words occur 20\% of the time.


\begin{minipage}{.5\linewidth}
\begin{table}[H]
\small
\caption{Candidate distribution for the word "\textit{nogensinde}".}
\begin{tabular}{c|lc}
\toprule
\textbf{Word} & \textbf{Error candidates} & \multicolumn{1}{l}{\textbf{Probability}} \\ \midrule
Nogensinde             & Nogen sinde    & 0.917                             \\
                       & Sinde          & 0.053                             \\
                       & Nogen          & 0.024                             \\
                       & Nogen sider    & 0.003                             \\
                       & Står           & 0.003                            \\ \bottomrule
\end{tabular}
\label{tab:nogensinde}
\end{table}
\end{minipage}%
\begin{minipage}{.5\linewidth}
\begin{table}[H]
\small
\caption{Candidate distribution for the word "\textit{lavet}".}
\begin{tabular}{c|lc}
\toprule
\textbf{Word} & \textbf{Error candidates} & \multicolumn{1}{l}{\textbf{Probability}} \\ \midrule
Lavet                  & Lave           & 0.409                             \\
                       & Lade           & 0.136                             \\
                       & Lavede         & 0.136                             \\
                       & Ladet          & 0.091                             \\
                       & \textit{Random}         & 0.227                        \\ \bottomrule    
\end{tabular}
\label{tab:lavet}
\end{table}
\end{minipage}%

\subsection{Topic model}
We train an LDA topic model on a Danish Wikipedia dataset consisting of 264,505 documents from The Danish Gigaword Corpus \cite{dagw} using the Gensim framework \cite{rehurek_lrec}. All documents are preprocessed by first performing word tokenization and removing punctuation and other special characters, including numbers. Next, we apply Part of speech (POS) filtering, keeping adjectives, nouns and verbs. Finally, we perform lemmatization, lowercase all letters, before finally vectorizing the documents into bag of word (BOW) representations. The BOW vocabulary may contain n-grams and is limited by removing uncommon n-grams that appear in less than ten documents and very common n-grams that occur in more than 90\% of the documents. We fix the LDA parameters $\alpha$ and $\eta$ as $\alpha = \frac{1}{|T|}$ and $\eta= 0.1$, and choose our remaining hyper-parameters by performing grid search optimizing for topic coherence, using the $U_{mass}$-coherence metric to choose the best model. We tune the following hyper-parameters: Topics $[10, 20, \underline{30}, 40, 50, 60, 70, 80, 90, 100]$, Vocabulary of BOW $[\underline{\rm Unigrams}, {\rm Unigrams} + {\rm Bigrams}]$ and Variational Bayes iterations $[5, 10, 15, \underline{20}]$. The underlined values are the values that yielded the best model in terms of $U_{mass}$ coherence score.

\subsection{Evaluation of Topic Robustness over Noisy Sources}
\label{section:evaluation}


For the experiments, we construct a baseline by computing the CS between topic vectors of two document sets $(S_1, S_2)$ as described in Section \ref{sec:3:1}. We then measure the robustness to noise as the average change in magnitude of cosine similarity scores after injecting noise to documents in $S_2$ at varying values of $\beta$ as described in Section~\ref{ss:noiseinjection}. We alter the noise substitution method between experiments to allow for a comparison between uniform and statistics based noise. 

We conduct two complementary experiments: 
\begin{enumerate}
    \item Testing for similarity between podcast descriptions ($S_1$) and noisy ASR transcripts ($S_2$) at varying levels of noise. This allows us to identify if the topic model is robust to transcription errors, and whether ASR transcripts contain enough information for the topic model to produce meaningful topic vectors.
    \item Testing for similarity between raw ASR transcripts ($S_1$) and noisy ASR transcripts ($S_2$) at varying levels of noise. This further investigates how robust the topic model is to transcription errors, under the assumption that the transcript provides meaningful topic vectors.
\end{enumerate}

For each experiment, we vary $\beta$ in $[0,1]$ with steps of $0.1$. To reduce variance, we repeat this procedure $50$ times and report the average and the standard error.

\section{Results and Discussion}
\label{s:resultsanddiscussion}

\begin{figure}[H]
  \begin{subfigure}[b]{0.49\textwidth}
   \captionsetup{width=.85\linewidth}
    \includegraphics[width=\textwidth]{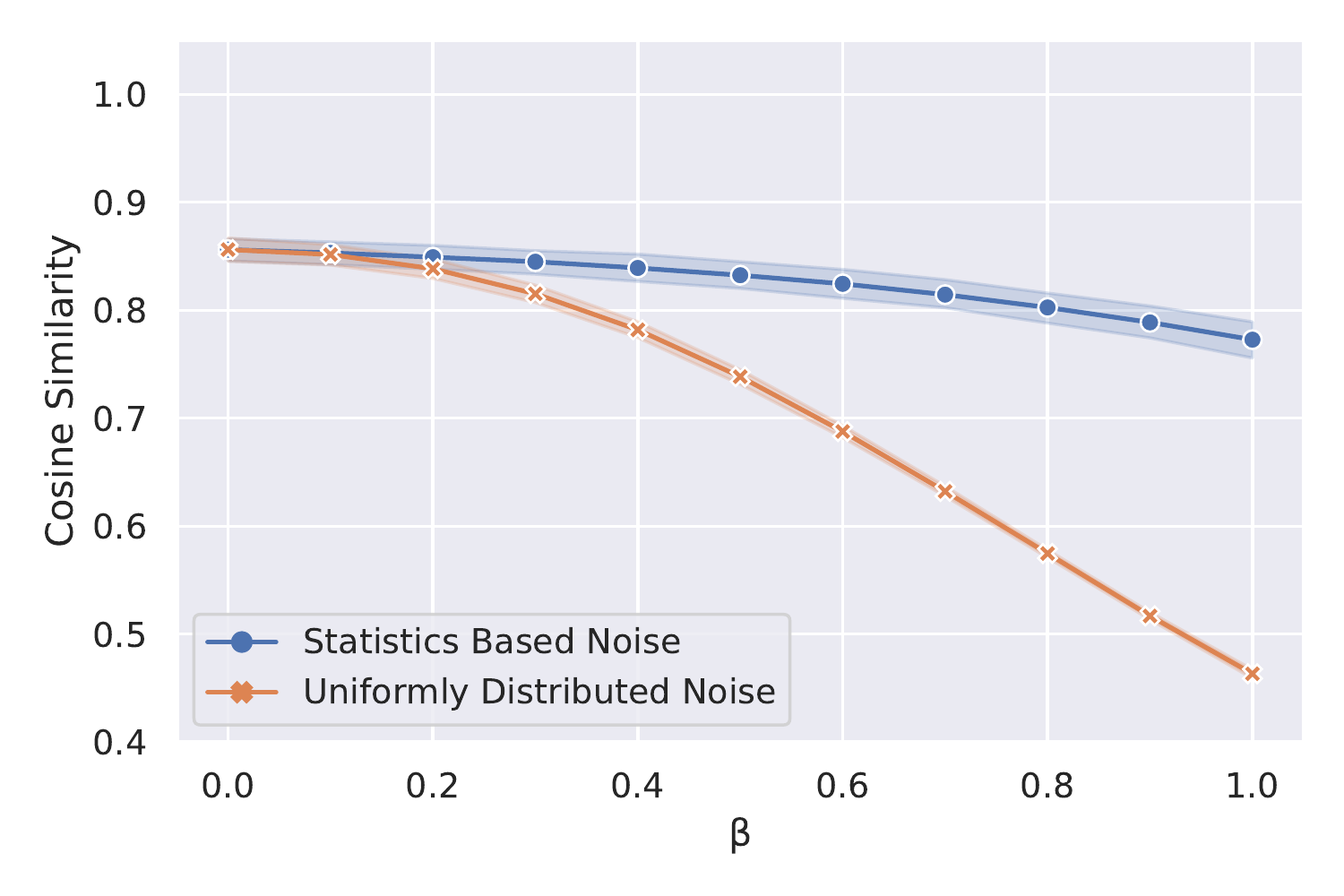}
    \subcaption{CS score between descriptions and noisy ASR transcripts at various levels of $\beta$.}
    \label{fig:results:1}
  \end{subfigure}
  \begin{subfigure}[b]{0.49\textwidth}
  \captionsetup{width=.85\linewidth}
    \includegraphics[width=\textwidth]{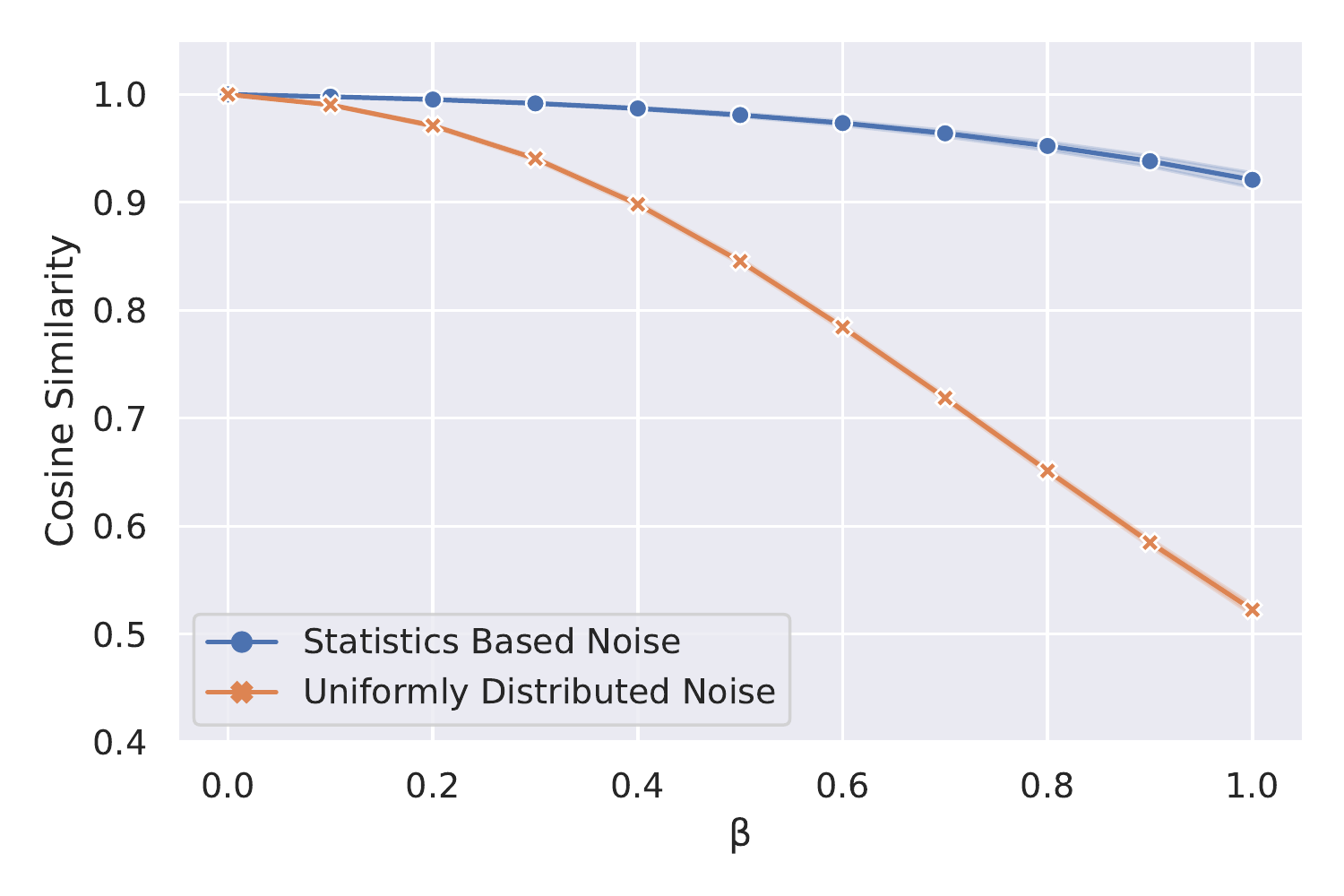}
    \subcaption{CS score between raw and noisy ASR transcripts at various levels of $\beta$.}
    \label{fig:results:2}
  \end{subfigure}
   \caption{Analysis of topic model robustness using cosine similarity, shaded area showing standard error of the mean.}
 \label{fig:results}
\end{figure}

The results of the two series of experiments are presented in Figure~\ref{fig:results:1} and Figure~\ref{fig:results:2}, respectively.

\begin{table}[!h]
\centering
\caption{The number of different podcast shows that are present in the cosine similarity deciles at $\beta=0$.}
\label{tab:uniqpodcasts}
\begin{tabular}{@{}ccccccccccccc@{}}
\toprule
 &\textbf{Decile} & 1 & 2 & 3 & 4 & 5 & 6 & 7 & 8 & 9 & 10&  \\ \midrule
\hspace{1pt} &\textbf{No. Unique podcasts} & 8 & 12 & 14 & 15 & 14 & 11 & 13 & 14 & 11 & 5& \hspace{1pt}\\ \bottomrule
\end{tabular}
\end{table}
As can be seen from Figure \ref{fig:results}, injecting noise in ASR transcripts has an effect on the CS for both noise distributions. We observe $45.9\%$ (0.85-0.46) and $49\%$ (1.00-0.51) relative decreases in CS when injecting uniform noise, while limited impact of  $8.8\%$ (0.85-0.775) and $9\%$ (1.00-0.91) is found for statistics based ASR noise. This is evidence that topic vectors are significantly more robust to statistics-based ASR noise than uniformly distributed noise. We also observe that the mean CS score between the descriptions and transcripts at $\beta = 0$ is $0.85$ which is significantly higher than what we observe as the lower bound, which can be roughly estimated to be $0.45$ (found at $\beta = 1$ after injecting uniform noise), which is evidence that ASR transcripts can be used to produce meaningful topic vectors. We suspect that the reason for the small change in topic distributions under the statistics based noise is mainly due to the type of errors produced by the ASR engine which, as seen in Table~\ref{tab:nogensinde} and Table~\ref{tab:lavet}, have a tendency to retain semantic meaning. Note that $\beta$ can be seen as an estimation of the WER in a podcast transcript (see Figure~\ref{fig:beta_vs_wer} in Appendix \ref{sec:appendix}), which suggests that even when an ASR engine produces a transcript with a high word error rate, the transcript will still be viable for topic modeling because the errors will have little impact on the topic distribution after LDA preprocessing.





When comparing transcripts and author descriptions for the baseline with $\beta=0$, the standard deviation has a value of 0.128. Furthermore, the difference between the minimum value of $0.500$ and the max value of $0.997$ is large. We investigate why this occurs by splitting the podcast episodes into deciles based on their cosine similarity and counting the number of unique podcast in each decile as seen in Table \ref{tab:uniqpodcasts}. The lowest amount of unique podcast shows are in the top and bottom deciles. Furthermore, in the 1st decile, 42 out of the 59 episodes comes from the same podcast "Sagen om Amagermanden" (note, it is also the podcast show that is the most represented) and in the 10th decile, 52 of the 58 episodes are from podcast show "Hvor er mit ansigt?". This result suggests that either the quality of the podcast descriptions or the performance of the ASR engine is very dependent on the specific podcast show. However, since all podcast shows in the dataset are single speaker with limited amount background music, we suspect that quality of the author descriptions is the primary reason why episodes originating from the same podcast show consistently perform poorly in some cases.

\section{Conclusion}
\label{s:conclusion}
In this work, we conducted experiments to evaluate the quality and robustness of topic representations produced by a general LDA topic model when exposed to noisy ASR transcripts in a low-resource language (Danish). More specifically, we investigated how injecting two different noise profiles to raw ASR transcripts influence the topic distributions across a podcast dataset at varying levels of noise. We created the dataset by leveraging an ASR system to obtain podcast transcripts. We choose to only include podcast episodes that had a high-quality author-written description as part of the meta-data in the dataset, to allow for comparison between the description and transcript topic vectors, relying on the assumption that a well-written description is very similar in terms of topic distribution to that of the episode content. To obtain topic vectors we trained a general-purpose LDA model on Danish Wikipedia data. We experimented with injecting two types of noise into the raw ASR transcripts, namely uniformly random noise and simulated ASR noise. We obtained similarity baselines by computing vector similarities of the raw transcripts with their respective descriptions and the raw transcripts with themselves at various levels of noise injection. We found that injecting random noise to the transcripts significantly influenced the CS score ($45.9\%$ and $49\%$ relative to baseline) whereas injecting simulated ASR noise only slightly influenced the CS score ($8.8\%$ and $9\%$ relative to baseline). Given our findings, we conclude that even when an ASR engine produces podcast transcripts at higher WERs, we can still obtain meaningful topic representations. We hypothesize that this is because even when an ASR engine produces erroneous text, the majority of the word-level errors will carry similar semantic meaning to the underlying truth. This encourages the use of ASR engines and transcriptions for podcast recommendations. Even in cases where the ASR engine is challenged by complex audio scenery, the transcripts could be sufficient to obtain topic representations that are relevant when representing the semantic content of a podcast episode. Furthermore, we also found implications that using ASR transcripts would be generally more robust than relying on human tinkered descriptions which are largely reliant on the individual content creators when it comes to performing topic modelling on podcast content. We conclude this based on the fact that episodes originating from the same podcast shows either consistently showed very high or very low topic overlap between their transcripts and their descriptions.

The ASR noise investigated in this paper was based on ASR errors from clean recordings without common podcast artifacts such as the overlap of speech and ambient sounds. In the future, it would be worth investigating whether topic models also are robust to ASR noise stemming from these types of errors. Furthermore, there are indications that other downstream tasks might not be as robust to noise as topic modeling, hence analyzing the robustness of other relevant downstream tasks to ASR noise remains an open research question.

\section{Acknowledgments}

We gratefully acknowledge support from Innovation Fund Denmark in the forms of their Innobooster and Innoexplorer grants (Grant numbers 0173-00670B and 0160-00023 respectively).

\bibliographystyle{ACM-Reference-Format}
\bibliography{bibliography}
\newpage
\appendix
\section{Additional Figures}
\label{sec:appendix}
\begin{figure}[!h]
    \centering
    \includegraphics[width=1\textwidth]{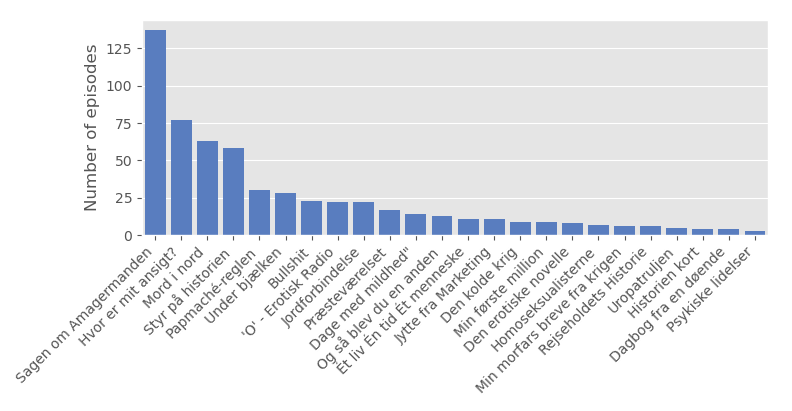}
    \caption{Distribution of episodes per podcast.}
    \label{fig:experiments:data}
\end{figure}

\begin{figure}[H]
\centering
\includegraphics[width=0.5\textwidth]{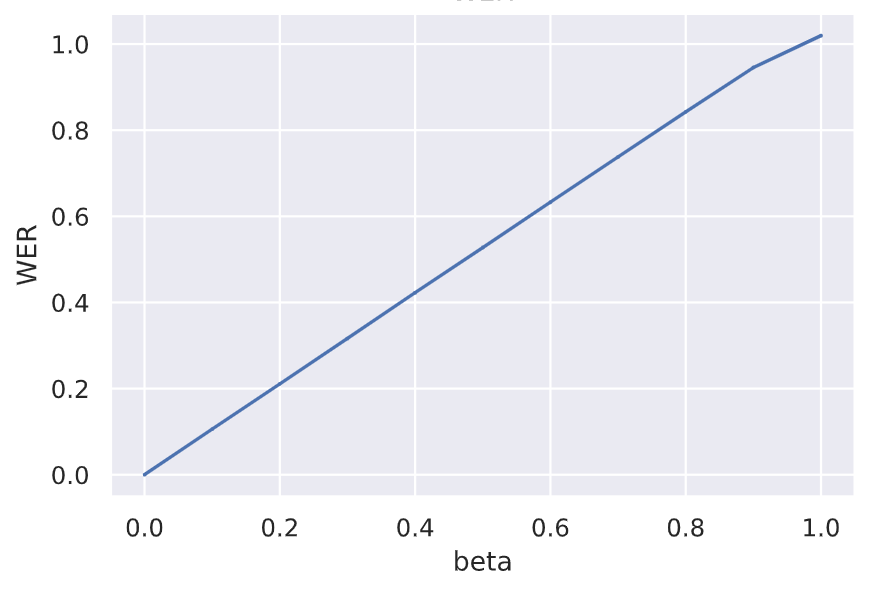}
\caption{Word Error Rate as function of $\beta$}
\label{fig:beta_vs_wer}
\end{figure}
\end{document}